# Linear effects of the nontraditional Coriolis terms on ITCZ-forced large-scale flow

Hing Ong[1]* and Paul E. Roundy[1]

[1]Department of Atmospheric and Environmental Sciences, University at Albany, State University of New York, Albany, New York

Corresponding author: Hing Ong (hwang2@albany.edu)

## Abstract

This paper promotes a measure to validate the hydrostatic approximation via scaling the nontraditional Coriolis term (NCT) in the zonal momentum equation. To demonstrate the scaling, this study simulates large-scale flow forced by a prescribed heat source mimicking the intertropical convergence zone (ITCZ) using a linearized forced-dissipative model. The model solves two similar equations between which the only difference is inclusion of NCTs. The equations are derived using the following approximations: anelastic, equatorial beta-plane, linearized, zonally symmetric, steady, and a constant dissipation coefficient. The large-scale flows simulated with and without NCTs are compared in terms of the meridional-vertical circulation, the zonal wind, and the potential temperature. Both results appear like the Hadley circulation. With the model parameters controlled, the results without NCTs minus with NCTs are linear biases due to omitting NCTs. The most prominent bias is a westerly wind bias in the ITCZ heating region that emerges because omitting NCTs prevents the associated westward acceleration when heating-induced vertical motion is present. The zonal wind bias divided by the zonal wind with NCTs is $0.120 \pm 0.007$ in terms of the westerly maximum and $0.0452 \pm 0.0005$ in terms of the root mean square (RMS) when the prescribed ITCZ mimics the observed ITCZ in May over the East Pacific. These normalized measures of the zonal wind bias increase with a narrower ITCZ or an ITCZ closer to the equator because of a weaker subtropical jet stream given the same vertical heating profile. This difference can be traced by a nondimensional parameter scaling the ratio of the NCT to the traditional Coriolis term. The scaling encourages restoring NCTs into global models.

## 1. Introduction

The nontraditional Coriolis terms (NCTs) are terms involving the meridional component of the planetary vorticity, $2\Omega \cos\vartheta$ ($\Omega$ and $\vartheta$ denote the rotation rate of Earth and latitude), in the zonal and vertical momentum equations. NCTs are omitted when the hydrostatic approximation is applied. The NCT in the vertical momentum equation is omitted when deriving the hydrostatic equation, and the NCT in the zonal momentum equation is omitted for dynamical consistency and energy conservation; the latter is called the traditional approximation (Eckart, 1960). The classical measure of validity of the hydrostatic approximation is based on scaling the terms omitted from the vertical momentum equation with respect to the terms left in the hydrostatic equation, and the classical scaling suggests that the hydrostatic approximation is valid for large-scale flow (e.g. Holton and Hakim, 2013; Vallis, 2017). However, the significance of the NCT in the zonal momentum equation affects the validity of the hydrostatic approximation because restoring one of the NCTs requires restoring the other for dynamical consistency and energy conservation. Scaling both NCTs in the tropical diabatic-forced large-scale flow, White and Bromley (1995) suggested that NCTs should be restored in global models. They proposed the quasi-hydrostatic equation set, which omits the vertical acceleration term and retains all the other terms.

Inspired by the scaling by White and Bromley (1995), Hayashi and Itoh (2012, hereafter HI12) simulated large-scale flow forced by a prescribed eastward-moving intraseasonal-oscillating heat source along the equator using a linearized forced-dissipative model. The model, originating from the quasi-hydrostatic equation set, solves two similar equations between which the only difference is whether they involve NCTs. The results without NCTs are consistent with previous studies (Gill, 1980; Schubert and Masarik, 2006), but the results with NCTs show an additional equivalent barotropic anticyclonic vorticity dipole maximized at the maximum heating level due to the planetary vorticity tilted by the meridional gradient of vertical motion, which contributes about 10% of the maximum vorticity. Then, Ong and Roundy (2019, *J. Atmos. Sci.*, under revision) analyzed the meridional Ertel potential vorticity (EPV) flux due to diabatic forcing coupled with NCT, $-2\Omega\dot{\theta}\cos\vartheta$ ($\dot{\theta}$ denotes the material temporal derivative of potential temperature), using reanalysis data. EPV is the dot product of absolute vorticity and potential temperature gradient divided by density. The significance of the NCT-diabatic EPV flux is assessed using the meridional mean-advective EPV flux in the tropics as a reference. In the ITCZ heating region, the NCT-diabatic EPV flux is a robust and considerable EPV discharger, where EPV discharging is defined as southward EPV flux and corresponds to EPV gradient reduction and balanced westward acceleration, and EPV charging means the opposite. Gerkema et al. (2008) reviewed various NCT-related phenomena. A common effect of NCTs is orienting the convection toward the rotational axis in giant-planet atmosphere, Earth ocean, and laboratory experiments.

This study focuses on ITCZ-forced large-scale flow. The coupled general circulation models (GCMs) have been unable to reasonably simulate the Earth's tropical mean climate and have a double-ITCZ bias (e.g. Neelin et al., 1992; Mechoso et al., 1995; Meehl et al., 2005; Xu et al., 2005; Lin, 2007; Song and G. J. Zhang, 2009; Bellucci et al., 2010; G. J. Zhang and Song, 2010; Hwang and Frierson, 2013; Li and Xie, 2014; X. Zhang et al., 2015). Many contributors to the double-ITCZ bias have been identified, including convective processes (e.g. Song and G. J. Zhang, 2009; Bellucci et al., 2010; G. J. Zhang and Song, 2010), radiative processes (e.g. Lin, 2007; Hwang and Frierson, 2013; Li and Xie, 2014), atmosphere-ocean interactions (e.g. Neelin et al., 1992; Lin, 2007; G. J. Zhang and Song, 2010), tropical-extratropical interactions (e.g. Hwang and Frierson, 2013), and Pacific-Atlantic interactions (e.g. Xu et al., 2005). Examining the ensemble means of two generations of Coupled Model Intercomparison Project (CMIP3, Meehl et al., 2007; CMIP5, Taylor et al., 2012) and contrasting them against observations, X. Zhang et al. (2015) showed that the spurious Southeast Pacific ITCZ, which doubles the ITCZ more often than nature, collocates with spurious near-surface weakened-easterly winds. Focusing on one model, G. J. Zhang and Song (2010) suggested that the weakened-easterly winds leads to warm bias in the ocean under the spurious ITCZ, which further enhances the spurious ITCZ. G. J. Zhang and Song (2010) revised the closure of the convective parameterization to strengthen the easterly trade winds and mitigated the double-ITCZ bias. However, the biases shown by X. Zhang et al. (2015) persisted in two generations of GCM ensembles with diverse formulations of convective parameterization, and this diversity causes difficulty to apply the revision of the closure by G. J. Zhang and Song (2010) to every convective parameterization. This difficulty motivates a hypothesis that the weakened-easterly winds, i.e. a westerly wind bias found in two generations of GCMs, may be due to neglecting NCTs in the dynamical cores, which is more fundamental than problems arising from a single convective parameterization.

The rest of this paper is organized as follows. Section 2 introduces a new linearized forced-dissipative model that can switch NCTs on and off. Adopting many concepts from HI12, the new model is simpler owing to the zonally quasi-symmetric and quasi-steady nature of the ITCZ. Section 3 assesses linear biases due to omitting NCTs, which are the results without NCTs minus with NCTs with the model parameters controlled. A nondimensional parameter that scales the ratio of the NCT to the traditional Coriolis term in the zonal momentum equation is introduced to explain the sensitivity of the zonal wind bias to changes in ITCZ location and width. Section 4 presents a summary along with a discussion of possible indirect effects of omitting NCTs and their implications to the double-ITCZ bias.

## 2. Methods

Following HI12, this study applies the equatorial beta-plane approximation and linearizes the equation set around a motionless stratified reference state. Furthermore, this study uses an anelastic approximation formulated in Lipps and Hemler (1982), which enables vorticity dynamics to govern the dynamical system (Jung and Arakawa, 2008). Also, the idealized ITCZ is assumed to be zonally symmetric and steady, so terms involving zonal or temporal derivatives are omitted. These approximations yield the following:

$$\alpha \theta' + \frac{d\tilde{\theta}}{dz} w = \frac{\tilde{\theta}}{c_p \tilde{T}} \dot{Q}, \tag{1a}$$

$$\alpha u - \beta y v + \boxed{2\Omega w} = 0, \tag{1b}$$

$$\alpha v + \beta y u + \frac{\partial}{\partial y}(c_p \tilde{\theta} \Pi') = 0, \tag{1c}$$

$$\alpha w \boxed{-2\Omega u} + \frac{\partial}{\partial z}(c_p \tilde{\theta} \Pi') - \frac{g}{\tilde{\theta}} \theta' = 0, \tag{1d}$$

$$\frac{\partial}{\partial y}(\tilde{\rho} v) + \frac{\partial}{\partial z}(\tilde{\rho} w) = 0. \tag{1e}$$

The variables are defined as follows: $\theta'$, perturbation potential temperature; $\Pi'$, perturbation Exner function; $u$, zonal velocity; $v$, meridional velocity; and $w$, vertical velocity. The parameters are defined as follows: $\tilde{\theta}$, reference potential temperature; $\tilde{T}$, reference temperature; $\tilde{\rho}$, reference density; $\dot{Q}$, heating rate; $\alpha$, dissipation coefficient; $c_p$ = 1004.5 J K$^{-1}$ kg$^{-1}$, heat capacity at constant pressure; $g$ = 9.81 m s$^{-2}$, acceleration due to gravity; $\Omega$ = 7.292 × 10$^{-5}$ s$^{-1}$, rotation rate of Earth; and $\beta \equiv df/dy$, where $f \equiv 2\Omega \sin \vartheta$. On the equatorial beta-plane, terms with explicit $2\Omega$ are NCTs (boxed) while terms with $\beta y$ are traditional Coriolis terms. Equations 1 originate from the nontraditional nonhydrostatic equation set for completeness, so the term $\alpha w$ is present in Equation 1d.

Define mass stream function, $\Psi$, so that $\tilde{\rho} v \equiv -\partial \Psi/\partial z$ and $\tilde{\rho} w \equiv \partial \Psi/\partial y$. Then, assuming constant $\alpha$ like in Gill (1980), a diagnostic equation for $\Psi$ can be derived;

$$(\alpha^2 + \boxed{4\Omega^2} + N^2)\frac{\partial^2 \Psi}{\partial y^2} + \boxed{2(2\Omega\beta y)}\frac{\partial^2 \Psi}{\partial y \partial z} + (\alpha^2 + \beta^2 y^2)\frac{\partial^2 \Psi}{\partial z^2}$$
$$+ \boxed{\left(\frac{2\Omega\beta y}{H}\right)}\frac{\partial \Psi}{\partial y} + \left(\frac{\alpha^2}{H} + \frac{\beta^2 y^2}{H} + \boxed{2\Omega\beta}\right)\frac{\partial \Psi}{\partial z} = \frac{\tilde{\rho}g}{c_p\tilde{T}}\frac{\partial \dot{Q}}{\partial y}, \quad (2)$$

where $N \equiv g \, \mathrm{d} \ln \tilde{\theta}/\mathrm{d}z$ and $1/H \equiv -\mathrm{d} \ln \tilde{\rho}/\mathrm{d}z$, i.e. buoyancy frequency and inverse density scale height. Detailed derivation from Equations 1 to Equation 2 can be found in Appendix A. The term $\alpha^2$ in the coefficient $(\alpha^2 + \boxed{4\Omega^2} + N^2)$ originates from the term $\alpha w$ in Equation 1d, so the inclusion of the vertical acceleration term is insignificant to the system if $\alpha$ is reasonably small. With a $\Psi$ field, $v$ and $w$ can be calculated using the definition, and $\theta'$ as well as $u$ can be calculated by rearranging Equations 1a and 1b;

$$\theta' = \frac{1}{\alpha}\left(\frac{\tilde{\theta}}{c_p\tilde{T}}\dot{Q} - \frac{\mathrm{d}\tilde{\theta}}{\mathrm{d}z}w\right), \quad (3a)$$

$$u = \frac{1}{\alpha}(\beta y v \boxed{-2\Omega w}). \quad (3b)$$

Therefore, the model first solves Equation 2 for $\Psi$. The results with and without NCTs are computed using the same numerical solver, and the only difference is that the boxed terms in Equations 2 and 3 are set to zero for the calculation without NCTs. Given the parameters used in this study, Equation 2 is elliptic. The model solves the equation iteratively using the multigrid method. The domain is rectangular; $y \in [-6400 \text{ km}, 6400 \text{ km}]$ and $z \in [0 \text{ km}, 32 \text{ km}]$. The boundary conditions are Dirichlet where $\Psi = 0$; that is, flow normal to the boundary is not allowed. The reference temperature profile is set by the following: $\tilde{T} = 300$ K at $z = 0$, $d\tilde{T}/dz = -6.5$ K km$^{-1}$ for the troposphere, $d\tilde{T}/dz = 2.6$ K km$^{-1}$ for the stratosphere, and the tropopause height is 16 km. With the reference surface pressure set to 101325 Pa, the reference pressure profile is derived from the reference temperature using the hydrostatic relation. The reference profiles of the other thermodynamic parameters are derived from the temperature and the pressure. The meridional and vertical grid spacings are 100 km and 0.5 km. The tolerance of numerical errors is 1/32768 of the right-hand side of Equation 2 in terms of the Euclidean norm.

The ITCZ-like heating rate is zero in the stratosphere while Gaussian in the meridional and squared sine in the vertical spanning the troposphere with an exponential vertical weighting;

$$\dot{Q}(y,z) = \dot{Q}_{\max}e^{\frac{-(y-\mu)^2}{2\sigma^2}}\sin^2\left(\frac{\pi z}{16 \text{ km}}\right)e^{\frac{\gamma z}{2H}} - R(z), \text{ for } z \in [0 \text{ km}, 16 \text{ km}]. \quad (4)$$

Hereafter, four forcing parameters are defined as follows: ITCZ maximum latent heating rate, $\dot{Q}_{\max}$; ITCZ location, $\mu$; ITCZ width, $4\sigma$; and $\gamma$, vertical weighting parameter. $R(z)$ is a background radiative cooling rate that adjusts the meridional mean of $\dot{Q}(y,z)$ to zero. Along with the dissipation coefficient, $\alpha$, there are in total five dimensions determining the forced-dissipative parameter space. However, changes in $\dot{Q}_{\max}$ and $\alpha$ are not worth exploring because the variables either not change or change proportionally. The magnitude of every variable changes proportionally with $\dot{Q}_{\max}$ because $\dot{Q}_{\max}$ enters the dynamical system only through the right-hand side of Equation 2. $\alpha$ seems to have many entries to the dynamical system, but the only considerable entries are the right-hand sides of Equation 3. $\Psi$ and thus $v$ as well as $w$ are not sensitive to changes in $\alpha$ unless $\alpha$ becomes larger than 1/2 day$^{-1}$, which is unreasonable. Hence, $\theta'$ and $u$ are inversely proportional to $\alpha$. These proportional changes with the parameters are trivial because they make no difference when the results are normalized. The vertical profile of the heating is treated as a source of uncertainties in this study. Over the East Pacific, the height of the heating maximum varies seasonally and ranges roughly from 3500 m to 8000 m, and May is around a period of the transition (Huaman and Schumacher, 2018). Accordingly, $\gamma = 0, –4, –8$ are tested to cover the range of uncertainties while $\gamma = 0$ is used for figure demonstration. Thus, this study explores a parameter space with two dimensions, the ITCZ location and the ITCZ width. A control simulation is chosen for demonstration purposes. In the control simulation, the prescribed ITCZ is set to mimic the observed ITCZ in May over the East Pacific (Figure 1); the ITCZ location and the ITCZ width are 600 km and 1000 km, and $\dot{Q}_{max}$ is adjusted so that the maximum vertically integrated heating rate yields 9 mm day$^{-1}$ of precipitation. The annual mean is not suitable for the demonstration because the ITCZ location varies seasonally. In May, the observed ITCZ transforms from double to single over the East Pacific, but the simulated ITCZ in many GCMs remains double (Bellucci et al., 2010). The observed ITCZ is based on GPCP Version 2.3 Monthly Analysis Product (Mesoscale Atmospheric Processes Branch and Earth System Science Interdisciplinary Center, 2018). The process of choosing $\alpha$ in the control simulation is iterative. $\alpha = 7.292 \times 10^{-7}$ s$^{-1} \cong 1/16$ day$^{-1}$ yields a reasonable westerly maximum. Also, this choice of $\alpha$ is comparable to the Newtonian cooling rate for studying idealized Hadley circulation (e.g. Held and Hou, 1980). For the parameter space, the ITCZ location spans from 0 km to 1600 km, and the ITCZ width spans from 400 km to 1600 km, both incrementing by 100 km. All the experiments are conducted with and without NCTs, and with three types of vertical heating profile. Consequently, the total number of simulations is $17 \times 13 \times 2 \times 3 = 1326$. MATLAB scripts implementing this model are available from: https://github.com/HingOng/NCT-ITCZ.

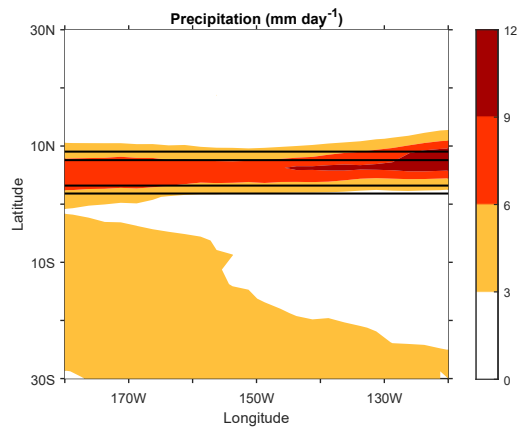

**Figure 1**. The horizontal distributions of the prescribed precipitation in the control simulation (contours) and the observed mean precipitation in May from 1979 to 2018 (shadings). The four contours denote 3, 6, 6, and 3 mm day$^{-1}$ from south to north.

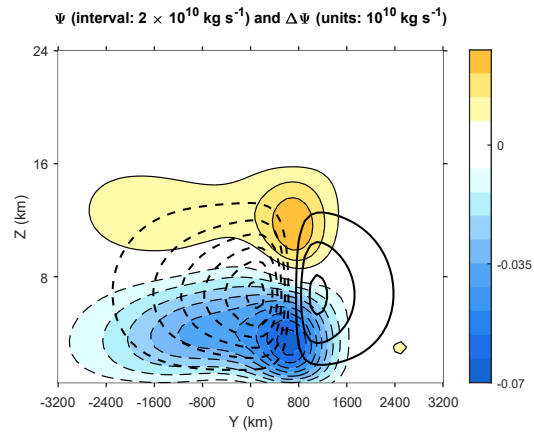

**Figure 2**. The meridional-vertical distributions of the mass stream function with NCTs (thick contours) and the differential mass stream function without NCTs minus with NCTs (thin contours with shadings) in the control simulation. The solid and dash contours denote positive and negative values. The zero contour is omitted.

## 3. Results and Scaling

Sanity checks on the mass stream function in the control simulation are performed because every other variable is calculated from it. First, the mass stream function with NCTs (Figure 2) appears like the Hadley circulation. The minimum and maximum contours are $–12 \times 10^{10}$ kg s$^{-1}$ and $6 \times 10^{10}$ kg s$^{-1}$, the absolute values of which lie within the range of seasonal variability, from $2 \times 10^{10}$ kg s$^{-1}$ to $22 \times 10^{10}$ kg s$^{-1}$ (e.g. Hartmann, 2016); to facilitate the comparison, Figure 2 depicts $\Psi$ multiplied by the circumference of Earth. Zonally asymmetric processes strengthen the Hadley circulation (e.g. Walker and Schneider, 2005), so the results are expected to scale like the climatology in terms of the order of magnitude but not to fit the climatology. Then, owing to the NCT-related coefficient in the $\partial y \partial z$ term in Equation 2, the circulation should be upright without NCTs and should be tilted upward and poleward with NCTs, consistent with Gerkema et al. (2008). Hence, omitting NCTs biases the circulation orientation toward upward and equatorward. In addition, the NCT-related coefficient in the $\partial y^2$ term in Equation 2 implies that omitting NCTs strengthens the circulation, compensating for the reduced coefficient. The mass stream function without NCTs minus with NCTs (Figure 2) is subtle in comparison to the total; the minimum and maximum contours are $-7 \times 10^8$ kg s$^{-1}$ and $2.8 \times 10^8$ kg s$^{-1}$. Yet the difference confirms that the model conforms to the expectations. In Figure 2, the minimum of the total is located upward and equatorward of the minimum of the difference, and the maxima are located similarly, indicating the circulation orientation bias. Also, the cross-equatorial circulation strengthens due to omission of NCTs.

Omitting NCTs causes a westerly wind bias (Figure 3). The westerly bias field is proportional to the heating rate field because heating-induced upward motion (Figure 4) yields westward acceleration via the NCT. The westerly bias also corresponds to quadrupole potential temperature biases (Figure 5) in thermal-wind balance; southward temperature gradient with westerly shear is in the lower troposphere, and the opposite is in the upper troposphere. The balanced features are consistent with the paradigms of both planetary vorticity tilting (HI12) and EPV charging (Ong and Roundy, 2019, *J. Atmos. Sci.*, under revision). The planetary vorticity tilting paradigm predicted an equivalent barotropic vorticity dipole with negative vorticity to the north of the heating maximum and positive to the south (Figure 3). The EPV charging paradigm predicted that NCT-coupled diabatic heating discharges (southward flux) EPV, so omitting NCTs makes the ITCZ a biasing EPV charger, which increases the meridional EPV gradient around the ITCZ; a positive vorticity bias (Figure 3) with enhanced stratification (Figure 5) is in the north, and the opposite is in the south. Considering adiabatic processes, the quadrupole potential temperature biases can also be explained by the quadrupole vertical motion biases (Figure 5), corresponding to the circulation orientation bias. In addition, adiabatic processes explain the cold bias in the ITCZ heating region and the warm bias in the other hemisphere (Figure 5), corresponding to the strengthened cross-equatorial circulation due to omitting NCTs. Except near the equator, the potential temperature biases overshoot into the stratosphere with a poleward tilt, which is, again, consistent with Gerkema et al. (2008).

Two normalized measures of the zonal wind bias are introduced. Since the maximum westerly wind in the subtropical jet stream (Figure 3) is not affected by omitting NCTs, the first normalized measure is the maximum westerly bias due to omitting NCTs divided by the maximum westerly wind with NCTs. Similarly, the second normalized measure is the ratio in terms of root mean square (RMS). In the control simulation, the maximum-based normalized zonal wind bias is $0.120 \pm 0.007$; the uncertainty estimate takes different vertical heating profiles into account. Similarly, the RMS-based bias is $0.0452 \pm 0.0005$. Hereafter, the normalized zonal wind bias is maximum-based unless otherwise noted. A nondimensional parameter, $\hat{O}$, is introduced to explain the normalized zonal wind bias;

$$\hat{O} \equiv \frac{aW}{YV}. \tag{5}$$

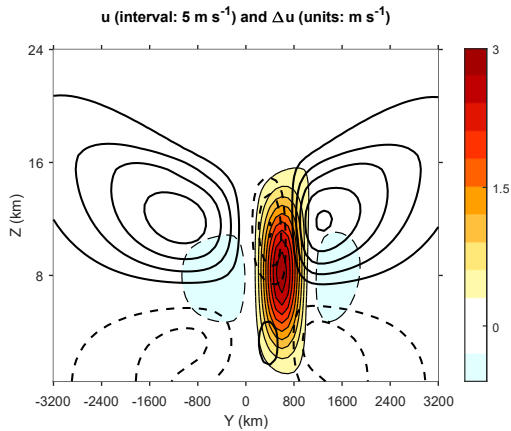

**Figure 3**. Like Figure 2 but for the zonal velocity with NCTs (thick contours) and the differential zonal velocity without NCTs minus with NCTs (thin contours with shadings).

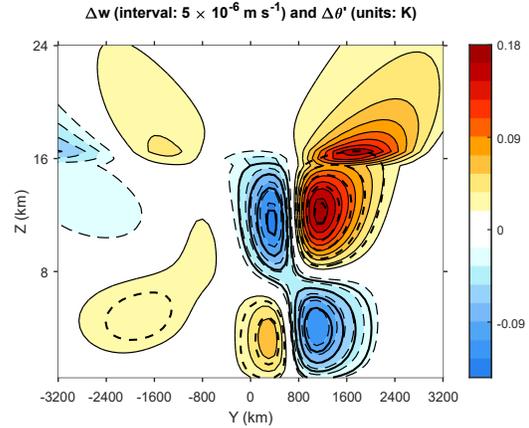

**Figure 5**. Like Figure 2 but for the differential vertical velocity without NCTs minus with NCTs (thick contours) and the differential potential temperature without NCTs minus with NCTs (thin contours with shadings).

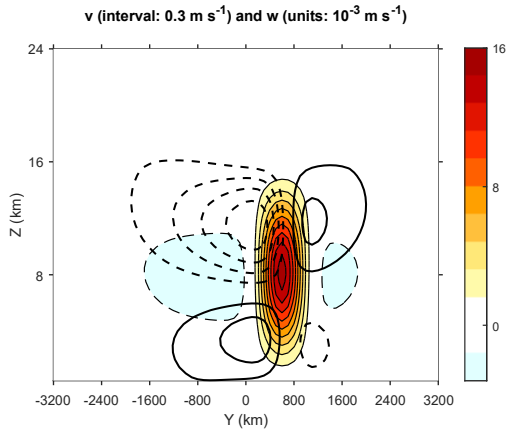

**Figure 4**. Like Figure 2 but for the meridional velocity with NCTs (thick contours) and the vertical velocity with NCTs (thin contours with shadings).

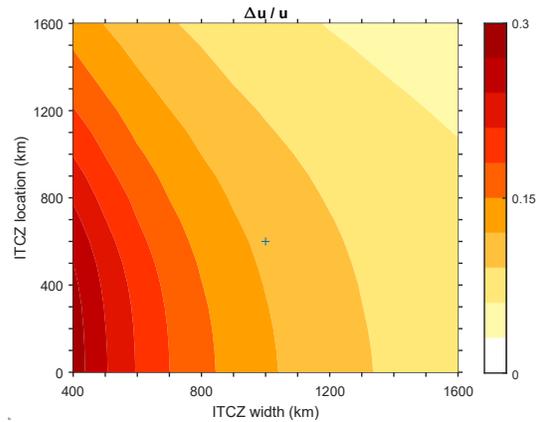

**Figure 6**. The distributions of the normalized zonal wind bias in a parameter space. The normalized zonal wind bias is calculated via dividing the maximum of the differential zonal velocity by the maximum of the zonal velocity with NCTs; e.g., the control simulation (Figure 3) is marked by the "+" sign. The ITCZ width covers an area within which 95% of the diabatic forcing occurs. The ITCZ location is defined using the equator as a reference.

$\hat{O}$ number is a measure of the ratio of the NCT to the traditional Coriolis term in the zonal momentum equation. The scaling variables are defined as follows: $V$, characteristic meridional velocity; $W$, characteristic vertical velocity; $Y$, characteristic distance from the equator; and $a$, radius of Earth. $a$ emerges because $\beta = 2\Omega/a$ on the equatorial beta-plane. Interpretation of Equation 5 can be straightforward; NCT is proportional to $W$, and the traditional Coriolis term is proportional to $V$ and the traditional Coriolis parameter, which is proportional to $Y$ on the equatorial beta-plane.

$\hat{O}$ number can be more useful when the $W/V$ ratio is related to the $D/L$ ratio, where $D$ and $L$ denote characteristic vertical depth and meridional length. In comparison to the $W/V$ ratio, the $D/L$ ratio is simpler to determine via observations and easier to manipulate in models. The mass continuity suggests that $W/V \sim D/L$ when the *Froude* number $\sim 1$ (e.g. Vallis, 2017). For zonally symmetric systems, like in this study, $W/V \sim D/L$ should be valid because vertical convergence of vertical mass flux must offset meridional divergence of meridional mass flux. Given $W$ and $D$, a larger $L$ requires a stronger $V$ to yield the same amount of meridional divergence, which leads to a stronger subtropical jet stream. This relation motivates another sanity check. Taking the troposphere depth as $D$ and the ITCZ width in the control simulation as $L$ yields $D/L = 0.016$. $W$ and $V$ in the control simulation are scaled using Figure 4. The peak vertical velocity lies between contours of $0.0128$ m s$^{-1}$ and $0.016$ m s$^{-1}$. The peak meridional velocity to the north of the ITCZ, where the maximum westerly wind locates, lies between contours of $0.6$ m s$^{-1}$ and $0.9$ m s$^{-1}$. Accordingly, $W/V$ lies between $0.014$ and $0.027$, which is comparable to $D/L$. Consequently,

$$\hat{O} \approx \frac{aD}{YL}. \tag{6}$$

While $a$, $D$, and $L$ are directly prescribed, $Y$ is not because the maximum westerly wind is characterized better by the location of the poleward branch of the Hadley circulation than the ITCZ location. A plausible measure of $Y$ is the ITCZ location plus a half of the ITCZ width, which is 1100 km in the control simulation. Accordingly, the $\hat{O}$ number is 0.093, which is comparable to the normalized zonal wind bias in the control simulation.

The normalized zonal wind bias changes in the ITCZ location-width parameter space (Figure 6). With the ITCZ narrowing, the normalized bias increases because the narrowing leads to a smaller $L$ and thus a larger $\hat{O}$ number. With the ITCZ approaching the equator, the normalized bias increases, but its rate of increase decreases. As for the reason, approaching the equator leads to a smaller $Y$ and thus a larger $\hat{O}$ number. Yet the approach also reduces the contribution of the ITCZ location to $Y$, and the ITCZ width becomes dominant. These results suggest that an increase in the normalized zonal wind bias in the parameter space can be explained by an increase in $\hat{O}$ number. With the bottom-heavy heating profile ($\gamma = -8$), the maximum westerly bias divided by the maximum subtropical jet strength appears like Figure 6 while differing by no more than 18% of the value shown in Figure 6. The RMS-based normalized zonal wind bias follows the above-mentioned patterns in the parameter space but is overall smaller than the maximum-based bias. The unnormalized westerly bias is mainly constrained by the maximum heating rate, so the strength of the subtropical jet due to poleward flow is the major factor determining the normalized zonal wind bias. HI12 suggested that the normalized zonal wind bias also increases with a meridionally narrower heating region, but the major determining factor is the strength of an easterly jet (Figure 3 of HI12) mainly due to equatorward flow (Figure 4 of HI12); since the traditional Coriolis terms also play the major role, $\hat{O}$ number may still apply.

### 4. Summary and Discussion

This paper promotes an alternative measure of validity of the hydrostatic approximation by scaling the nontraditional Coriolis term (NCT) in the zonal momentum equation instead of scaling the terms in the vertical momentum equation. A rationale for the alternative measure is that the hydrostatic approximation is affected by restoring the NCT in the zonal momentum equation while maintaining dynamical consistency and energy conservation. To demonstrate the scaling, this study simulates large-scale flow forced by a prescribed ITCZ-like heat source using a linearized forced-dissipative model. The simulations without NCTs are conducted using the same numerical solver as the simulations with NCTs but setting the NCT-involving terms to zero. The model equations are derived using the following approximations: anelastic, equatorial beta-plane, linearized, zonally symmetric, steady, and a constant dissipation coefficient.

Comparisons between the large-scale flow simulated with and without NCTs focus on the meridional-vertical circulation, the zonal wind, and the potential temperature. Hadley-like circulation occurs in both results. Linear biases due to omitting NCTs are calculated by subtracting the results of the simulation with NCTs from the results of the simulation without NCTs, with the model parameters held constant. The most prominent bias is a field of westerly bias proportional to the heating field that emerges because omitting NCTs prevents the associated westward acceleration when heating-induced vertical motion is present. The zonal wind bias is normalized by dividing the maximum westerly bias by the maximum westerly wind with NCTs or dividing the RMS of zonal wind bias by the RMS of zonal wind with NCTs. The maximum-based and RMS-based normalized zonal wind biases are $0.120 \pm 0.007$ and $0.0452 \pm 0.0005$ in the control simulation, where the prescribed ITCZ mimics the observed ITCZ in May over the East Pacific; the uncertainty estimate accounts for different vertical heating profiles. While changes in the heating rate or dissipation coefficient do not affect the normalized zonal wind biases, when an ITCZ narrows or approaches the equator, it enhances the normalized zonal wind biases via weakening the subtropical jet stream given the same vertical heating profile. To explain the sensitivity of the normalized zonal wind biases to changes in ITCZ geometry, a nondimensional parameter, $\hat{O}$, is introduced. The $\hat{O}$ number scales the ratio of the NCT to the traditional Coriolis term in the zonal momentum equation. The $\hat{O}$ number is 0.093 in the control simulation. A larger $\hat{O}$ number leads to larger normalized zonal wind biases and affects the validity of the hydrostatic approximation.

Using our model, the Hadley circulation and the mean zonal wind field can be reasonably simulated, and the temperature field is reasonable in spatial patterns but too extreme in magnitude. The large-scale temperature variation on a vertical level in the tropics is typically less than 1 K (e.g. Holton and Hakim, 2013; Vallis, 2017). However, in our model, the potential temperature difference between the maximum and the minimum reaches 10 K on the maximum heating level within 1600 km from the equator. We speculate that the rate of heat dissipation should increase with the magnitude of temperature anomaly so that extreme temperature anomalies are more prone to dissipation, but such a nonlinear process is beyond the scope of this study.

The results may have implications to tropical mean climate biases in GCMs yet not straightforwardly. If the GCMs had simulated accurately the diabatic forcing associated with the ITCZ, it would have had the linear biases due to omitting NCTs as shown in this study. However, precipitation in GCMs is not prescribed but coupled to dynamical variables, so those linear biases may bias the simulated precipitation indirectly. We hypothesize that the near-surface weakened-easterly wind, i.e. a westerly wind bias, which can lead to the spurious Southeast Pacific ITCZ (G. J. Zhang and Song, 2010; X. Zhang et al., 2015), may be due to neglecting NCTs in the dynamical cores. The results of this study suggest that an ITCZ-like heat source can cause a westerly bias in the heating region if NCTs are neglected. However, the results are limited by two caveats. First, after trade winds cross the equator, the easterlies remain easterly in observations but become westerly in the linearized model because horizontal advection is omitted; the spurious low-level westerly wind becomes more severe with a bottom-heavy heating profile. Second, the simulated westerly wind bias cannot reach the surface because boundary-layer processes are omitted. These two nonlinear effects are left for future studies.

The scaling encourages restoring NCTs into global models to improve the simulated tropical mean climate. Besides the mean climate, NCTs are important in other modeling aspects, including symmetric stability (e.g. Fruman and Shepherd, 2008; Itano and Maruyama, 2009) and wave dynamics (e.g. Kohma and Sato, 2013; Roundy and Janiga, 2012). White and Bromley (1995) suggested that restoring NCTs is more important than restoring the vertical acceleration term, which has been further supported by the present study and Hayashi and Itoh (2012) in terms of large-scale response to tropical diabatic heating. The development of the Met Office's Unified Model has followed this rationale (Davies et al., 2005). However, the development of many other models went on a different pathway. For example, DWD's ICOsahedral Non-hydrostatic model (ICON, Zängl et al., 2015), GFDL's Finite-Volume Cubed-Sphere model[1] (FV3), and NCAR's Model for Prediction Across Scales (MPAS, Skamarock et al., 2012) involved the vertical acceleration term but not NCTs when first developed. On the way to restore NCTs, ICON has accomplished (Borchert et al., 2018, under review), and FV3 is progressing (Hann-Ming Henry Juang, 27 Nov 2018, personal communication). We are starting to restore NCTs and deep atmospheric dynamics into MPAS.

**Appendix A**

To derive Equation 2 from Equations 1, first, substitute $v$ and $w$ with $\Psi$;

$$\alpha\theta' = \frac{\tilde{\theta}}{c_p \tilde{T}}\dot{Q} - \frac{d\tilde{\theta}}{dz}\frac{1}{\tilde{\rho}}\frac{\partial \Psi}{\partial y}, \tag{A1a}$$

$$\alpha u = -\beta y \frac{1}{\tilde{\rho}}\frac{\partial \Psi}{\partial z} \boxed{- 2\Omega \frac{1}{\tilde{\rho}}\frac{\partial \Psi}{\partial y}}, \tag{A1b}$$

$$\alpha \frac{1}{\tilde{\rho}}\frac{\partial \Psi}{\partial z} - \beta y u - \frac{\partial}{\partial y}\left(c_p \tilde{\theta} \Pi'\right) = 0, \tag{A1c}$$

$$\alpha \frac{1}{\tilde{\rho}}\frac{\partial \Psi}{\partial y} - 2\Omega u + \frac{\partial}{\partial z}\left(c_p \tilde{\theta} \Pi'\right) - \frac{g}{\tilde{\theta}}\theta' = 0. \tag{A1d}$$

Next, perform $\partial(A1c)/\partial z + \partial(A1d)/\partial y$ to derive a zonal vorticity equation;

$$\alpha \frac{1}{\tilde{\rho}}\frac{\partial^2 \Psi}{\partial y^2} + \alpha \frac{\partial}{\partial z}\left(\frac{1}{\tilde{\rho}}\frac{\partial \Psi}{\partial z}\right) \boxed{-2\Omega \frac{\partial u}{\partial y}} - \beta y \frac{\partial u}{\partial z} - \frac{g}{\tilde{\theta}}\frac{\partial \theta'}{\partial y} = 0. \tag{A2}$$

Then, multiply Equation A2 by $\alpha \tilde{\rho}$, and plug Equations A1a and A1b to eliminate $\theta'$ and $u$;

$$\alpha^2 \frac{\partial^2 \Psi}{\partial y^2} + \alpha^2 \tilde{\rho}\frac{\partial}{\partial z}\left(\frac{1}{\tilde{\rho}}\frac{\partial \Psi}{\partial z}\right) + \boxed{2\Omega \frac{\partial}{\partial y}\left(\beta y \frac{\partial \Psi}{\partial z} + 2\Omega \frac{\partial \Psi}{\partial y}\right)} + \beta y \tilde{\rho}\frac{\partial}{\partial z}\left(\beta y \frac{1}{\tilde{\rho}}\frac{\partial \Psi}{\partial z} + \boxed{2\Omega \frac{1}{\tilde{\rho}}\frac{\partial \Psi}{\partial y}}\right)$$
$$+ \frac{g}{\tilde{\theta}}\frac{\partial}{\partial y}\left(\frac{d\tilde{\theta}}{dz}\frac{\partial \Psi}{\partial y} - \frac{\tilde{\rho}\tilde{\theta}}{c_p \tilde{T}}\dot{Q}\right) = 0. \tag{A3}$$

Last, rearranging Equation A3 can yield Equation 2.

---

[1] In the final version, we mistakenly used "Finite-Volume model version 3" as the full name of FV3. This paper was composed during the 2018–19 United States federal government shutdown, so there was no reliable information source about FV3. We apologize for forgetting to correct this mistake before final publication.

**Acknowledgements**

This work was funded by National Science Foundation (grants AGS1757342, AGS1358214 and AGS1128779). We thank Dr. Justin Minder, Dr. Isaac Held, and two anonymous reviewers for the useful comments.